\newcount\pageformat\pageformat=2  
\newcount\pssize

\ifnum\pageformat=2
 \documentclass[aps,prb,twocolumn,floatfix,showpacs,amsmath,amssymb]{revtex4}
\else
 \documentclass[aps,prb,preprint,floatfix,showpacs,amsmath,amssymb]{revtex4}
\fi
\usepackage{graphicx}

\begin{document}
\title{Quantum Elastic Net and the Traveling Salesman Problem}

\author{B.\ F.\ Kostenko, J.\ Pribi\v{s}, M.\ Z.\ Yuriev}

\address{Joint Institute for Nuclear Research, Dubna,
141980 Moscow region, Russia \\}

\date{\today }

\begin{abstract}
Theory of computer calculations strongly depends  on the
nature of elements  the computer is made of. Quantum interference
allows to formulate the Shor factorization algorithm turned out to
be more effective than any one written for classical computers.
Similarly,  quantum wave packet reduction  allows to devise the
Grover search algorithm which outperforms  any classical one. In
the present paper we argue that the quantum incoherent tunneling
can be used for elaboration of new algorithms able to solve some
NP-hard problems, such as the Traveling Salesman Problem,
considered to be intractable in the classical theory of computer
computations.\end{abstract}

\pacs{ 03.67.Ac, 03.65.Xp, 03.65.Yz, 75.60.Ch }

\maketitle

\section{\label{sec1}Introduction}
Quantum parallel computations strongly differ  from parallel
calculi on the usual computers in a sense they use the same
physical processor for all parallel operations. For example, if
one wants to solve some property recognition problem one can
prepare an initial state in the form: $$ |v_0\rangle =
\frac{1}{\sqrt{2}} \sum_{\vec x} |\vec x, 0\rangle, $$ where $\vec
x$ describes a possible assignment to binary variables, $x_i$, of
the problem, $\vec x = (x_1, x_2, ... , x_n),$ $x_i = 0;1$, and
the last bit in the array $v_0$ is left to store a result of
calculations, $f(\vec x)$, which is also presented in the binary
form: $$f(\vec x)=1,$$ if the set $(x_1, x_2, ... , x_n)$ has a
required property, and $$f(\vec x)=0$$ otherwise. Calculations are
performed during time $t$ when the quantum computer evolves under
unitary evolution operator,
\begin{equation}\label{1}
|v_0\rangle  \rightarrow |v_t\rangle = U_t |v_0\rangle =
\frac{1}{\sqrt{2}} \sum_{\vec x} |\vec x, f(\vec x)\rangle .
\end{equation}
In that way the results of calculations for all possible
assignments of the variables $x_1, x_2, ... , x_n$   turn out to
be simultaneously evaluated and recorded on the same physical
equipment.

It is clear that this method of computation gives a great economy
of computer's time and memory. Unfortunately, the described
straightforward strategy is hampered by the following essential
reason. Quantum mechanics says that if you try now to read out the
results of calculations, you will succeed in observing only one of
them, corresponding to some particular assignment of $\vec x$, and
all other results will be irrevocably lost. Furthermore, the
typical state of affairs with hard combinatorial problems is that
results of calculations for most of assignments of $ \vec x$ are
equal to $0$, and there are very few of them (and namely they are
of interest!) which correspond to $f(\vec x)=1$. According to
quantum mechanics, for the vector $|v_t\rangle$ in form (\ref{1}),
any particular result of observation of $f(\vec x)$ will occur
with probability $P(\vec x)=1/2^n$. It is definitely forbidden to
pick out at will any particular result of
calculations on the stage of reading out
. You can amplify the contribution of a result you want to know in
the total wave packet only on the stage of the dynamical evolution
described, as in the above given example, by the unitary evolution
operator $U_t$, or in the general case, by more general operator
taking into account dissipation and/or measurement processes. In
other case, the property recognition problem can not be efficiently
resolved (is intractable) since the probability of observation of a
necessary result is exponentially suppressed as $P\sim 2^{-n}$.

Actually, there is no difference between calculations by quantum
computers and the stochastic (Monte Carlo) calculations by
classical computers, if one does not use some distinctive features
of quantum systems to increase the probability of the result you
want to obtain. For example, the {\it quantum interference} was
used by P. Shor to determine a period of unknown function in his
famous factorization algorithm \cite{Shor}. This application of
quantum mechanics is similar to usage of the interference for
determination of a period of a crystal by Bragg's scattering
method. The Grover search algorithm increases the probability of
the desired event with the help of the quantum {\it wave  packet
reduction} \cite{Grover}. For the above considered example of
property recognition problem, it is necessary to obtain after
dynamical evolution a wave packet at which the components with
$f(\vec x)=1$ are present with probabilities much bigger than the
components with $f(\vec x)=0$.

In this paper we give some plausible arguments in favor of
possibility of application of the quantum {\it incoherent
tunneling} for solving some NP-hard combinatorial problems.

\section{\label{sec2}ANALOG VERSUS DIGITAL COMPUTATIONS}
The main difference between analog and digital computers consists
in a manner of representation of  numbers in them. In digital
computers, the digital (usually binary) encoding of numbers is
used, e.g., the value of $x$ may be represented as $100101$. In
analog computers, different physical values, such as electrical
currents, angles of rotation of gear wheels, etc, are
representatives of mathematical values. Because of this the
digital computers are much more compact and their size, $L$,
scales with the quantity, $n$, of bits used for representation of
the numbers  as $ L \sim n.$ Enclosing such numbers in analog
computers requires much more large physical storage device, $ L
\sim e^n,$ and, therefore, corresponding resources, such as
energy, forces etc necessary for operations with the numbers are
also exponentially large in this case.

One more difference following also from the manner of number
representation concerns the property of  universality: digital
computers are ordinarily  universal, analog computers are designed
for solving some special problems only. For each particular
digital computer, universality should be proved, i.e. every
arithmetic, logical, and other operations should be constructed of
the {\it basic operations} with digits.

Usually merely computations that do not use resources that grow
exponentially   are of interest (so-called {\it efficient}
computations). For them the {\it Strong Church's Thesis} was
formulated \cite{Vergis}: Any finite analog computer can be
simulated {\it efficiently} by a digital computer, in a sense that
the time required by the digital computer to simulate the analog
computer is bounded by a polynomial function of the resources used
by the analog computer. This Thesis resulted from a great
experience gained during elaboration of classical computers and
has a clear sense: every classical physical process   can be
efficiently simulated with the help of the digital computers.

Due to a great progress in development of digital computer
techniques, analog computers are used now extremely seldom.
Nevertheless,  they survived in a virtual form inside digital
computers as  analog algorithms and turned into powerful heuristic
methods for solution of optimization problems. The analog
algorithms inherited from their natural ancestors the
correspondence of computational schemes to some real physical
processes in nature. Thus, the {\it steepest descent} method is
based on gradient equations often used for far-from-equilibrium
physical system description \cite{Rabinovich}. It is used to find
local minima. {\it Simulated annealing} describes the opposite
situation when a physical system is  very close to the equilibrium
state at each moment of its slow dynamical evolution. It is an
example of analog algorithms for minimization of a function called
energy which imitates the process of physical annealing known in
practice of crystal growing. It is experimentally discovered and
theoretically understood that molten matter, to be subjected to
slowly cooling, transforms into a crystalline state corresponding
to the minimum of its free energy. Simulated annealing was the
only method (in addition, of course, to the honest looking over
all variants) which gave hope  to find the global minimum of
combinatorial problems. However, rigorous consideration showed
that in fact the global minimum can never be reached
\cite{Azencott}, because simulated annealing with cooling schedule
$$T_1, \; T_2, \; ..., \;  T_n \rightarrow 0,  $$ requires an
exponential computation time: $$t_n \sim \exp (A/T_n), \qquad A=
const. $$ Main computation time losses take place at the lowest
temperature reached during calculations and the minimum becomes
more and more inaccessible at each step of cooling. Even a very
simple minimization problem corresponding to a biased double well
energy function shown in Fig.1 requires the cooling schedule $$T_n
\sim D/\ln \; n $$ and an exponential computational time, $t \sim
\exp n$, and, therefore, is intractable in the frame of this
algorithm.

\begin{figure}[!ht]
\includegraphics[width=6cm]{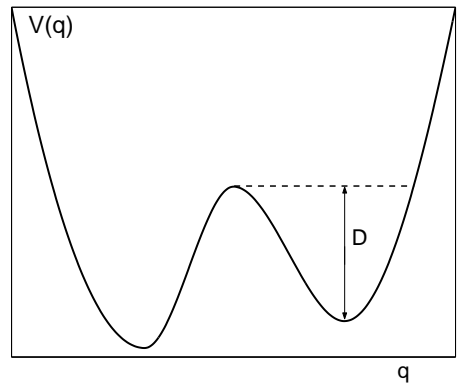}
\caption{\small Minimization problem corresponding to a biased
double well energy function requires the cooling schedule $T_n
\sim D/\ln  n $ and an exponential computational time in the frame
of the simulated annealing algorithm.}\label{fig1}
\end{figure}

\begin{figure}[!ht]
\includegraphics[width=6cm]{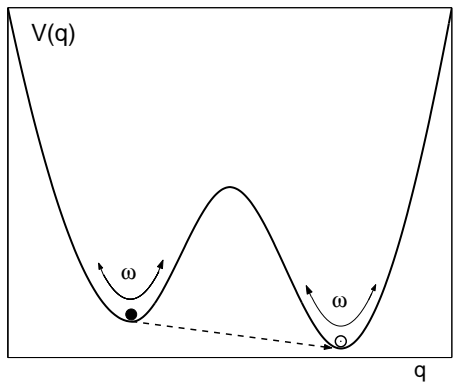}
\caption{\small Tunneling with dissipation allows particle to
penetrate through the potential barrier in a finite space of time
and then to stay at the point of the global minimum. }\label{fig2}
\end{figure}

It appears at first sight,  one can somehow accelerate cooling  if
does not require reaching the energy minimum on certainty. Then,
after several reiteration, the position of the global minimum
could  be selected among all local minima found. But this strategy
is, of course, wrong because no predictions about possibility to
recognize the global minimum could be done if the true cooling
schedule is broken.


Description of  quantum systems often requires consideration of
exponentially large number of different variants, e.g.  a
numerical realization of the path integral approach, and in
particular for this purpose R.~Feynman suggested quantum computers
would be especially effective \cite{Feynman}. For  quantum
computations no statements similar to the Strong Church's Thesis
were formulated so far and, therefore, both analog and digital
approaches are still of equal interest. Below, we discuss an
approaches to solution of the Traveling Salesman Problem by a
quantum analog computing machine.

\section{\label{sec3}INCOHERENT TUNNELING}

In this section plausible reasoning are given in favor of
incoherent tunneling as an effective remedy for solution of
minimization problems on the  stage of low temperatures when the
simulated annealing algorithm is inoperative. To this goal let us
consider again a biased double well potential shown in Fig.2 and a
particle localized initially in the left local minimum at point
$q_0$. Due to quantum tunneling effect, the particle can penetrate
the potential barrier and fall into the global minimum at $q_1$.
When the motion of the particle is accompanied by dissipation,
there is a chance that the particle will stay at the global
minimum and, hence, an optimization problem will be resolved.

A probable picture sketched above may be rigorously grounded. In
the case when  thermal energy, $k_B T$, is much less than energy,
$\hbar \omega$, of classical oscillations  at the positions of
minima and the latter is, in its turn,  much than the height of
the potential barrier, $V_{bar}$, between the wells, $$ k_B T \ll
\hbar \omega \ll V_{bar} , $$ the process of tunneling with
dissipation can be described on the base of multi-instanton
calculations in the framework of the imaginary time functional
integral approach \cite{Weiss}. For temperature close to zero, the
tunneling rate from $q_0$ to $q_1$ is given by an expression: $$
\gamma (q_0 \rightarrow q_1) = \frac{\pi}{2}
\frac{\triangle^2}{\omega} \frac{1}{\Gamma (2 \alpha)} \left(
\frac{\sigma}{\omega}\right)^{2\alpha -1} , $$ while the inverse
transitions are  suppressed, $$ \gamma (q_1 \rightarrow q_0) = 0.
$$ Here $\alpha = \eta d^2 / 2\pi \hbar$ is dimensionless damping
coefficient, $d=q_1 - q_0$ is the tunneling length, $\eta$
describes the force of friction, $F= - \eta dq/dt,$ and
$\triangle$ is the tunnel matrix of undamped and unbiased system.
The time of transition to the minimum, $\tau$, is {\it finite}
now: $$ \tau \sim \frac{\hbar}{\gamma} . $$ Thus, at least for the
double well, tunneling with dissipation, or incoherent tunneling,
can be effectively applied for solving the minimization problem in
the vicinity  of zero temperatures.

\begin{figure}[!ht]
\begin{center}\vspace{-.6cm}
\includegraphics[width=6cm,angle=-90]{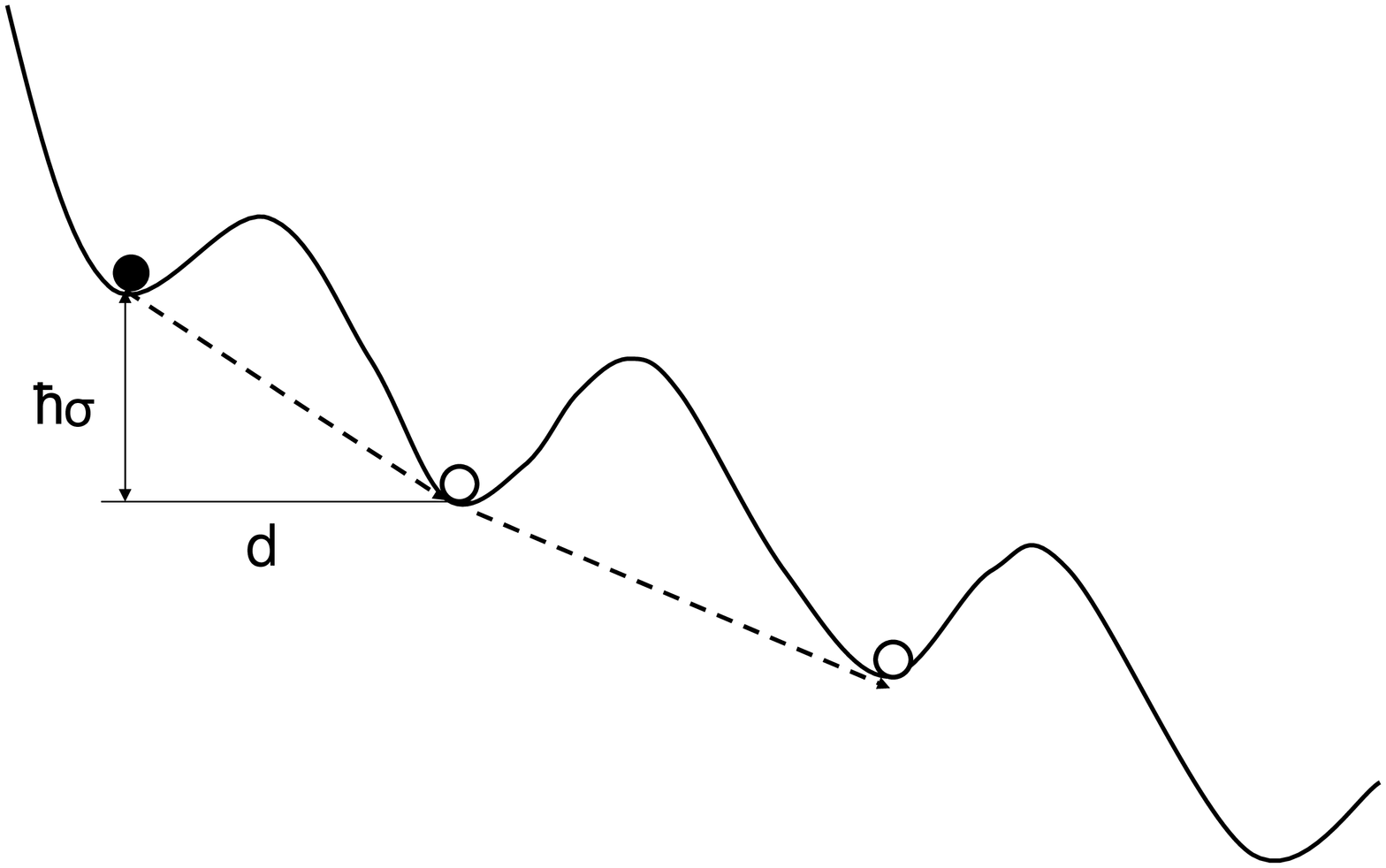}
\end{center}\vspace{-1cm}
\caption{\small Particle moves down the tilted periodic potential
hill with a constant speed.}\label{fig3}
\end{figure}

In the case of a tilted periodic potential, displayed  in Fig.3,
it is also possible  to fulfil calculations explicitly and show
that the time of transition from one local minimum to another is
finite. An average position of the particle may be expressed as
\cite{Weiss} $$ \langle q(t) \rangle = v \; t,$$ where $v$ is mean
drift velocity, $$ v=\frac{d  \hbar}{\gamma} \; \tanh (\beta \hbar
\sigma/2) , \qquad \beta = 1/k_B T .$$ Therefore, the time of
minimization is proportional now to the sum of the tunneling time
through all potential barriers on the way to the global minimum.
It is naturally to suggest that a similar picture remains in the
general case when the slope of the "hill" is variable. Thus, the
main question left to be discussed is how many local minima
actually are on the way to the global minimum. In the case of the
exponential number of them the problem would remain open.

The practice of solving optimization problems shows that usually
there are much less of  energy local minima than all possible (non
minimal) values of energy. For example, number of local minima of
the energy function was thoroughly investigated for the Hopfield
Network \cite{Hopfield} which was suggested as a mathematical
model of the associative memory in human brain. Here each local
minimum $H(\vec x)$ is used to retain  an information about $\vec
x$ (called a pattern) since for a small deviation from $\vec x$,
$\vec x \rightarrow \vec x  + \vec {\delta x}$, the system returns
back into $\vec x$ if it is navigated with the help of the
gradient equations. It was shown that for random created patterns
the number of local minima $M$ can be estimated as $$ M= 0.138 \;
n,$$ where $n$ is a number of artificial neuron cells
\cite{Hertz}. Thus $M$ is only a {\it linear} function of  the
size, $n$, of the input. A similar result was obtained for an
associative memory based on $Q$-state Potts-glass with biased
patterns \cite{Bolle}.

Further increase of the number of stored pattern up to $\alpha =1$
converts the Hopfield Network into a spin-glass \cite{Kinzel}. The
problem of finding the ground state in a spin-glass was studied in
\cite{Barahona}. This problem was shown to belong to the class of
NP-hard problems both for three-dimensional case, and for
two-dimensional lattice within a magnetic field. Infinite-ranged
models of spin-glasses were considered in \cite{Kirkpatrick}.
Numerical experiments have shown  the total number of local minima
increases as some {\it small power} of $n$, rather than $\exp
(n)$.

A drastic decrease of the number of local minima with different
energies in the vicinity of the global one, for any system in the
thermodynamic limit, follows also from the Nernst theorem for
entropy, $$ \lim_{T \rightarrow 0} S(T) = 0, $$ if there are no
gaps of function $S(T)$ near $T=0$\footnote{In principle, the gaps
are possible in the case of presence of phase transitions of the
first kind at $T$ close to zero.}.

It is a commonly accepted method to demonstrate an efficiency of a
new quantum algorithm with the help of a classical computer
simulation of corresponding quantum calculations. But in our case
there is no need to do so, because this has been {\it already} done.
Actually,  some kind of incoherent tunneling was used efficiently
many times as an optimizing procedure during simulated annealing.
For instance, in \cite{Kirkpatrick} not only neighboring (in the
Hamming sense) configuration of spin were checked during
minimization, but also states which were 2, 3 and 4 steps away. When
energy of a trial state was found to be less than energy of the
current state the system was transmitted to the trial state. This
corresponds to the under-barrier transitions through the potential
barriers with 1, 2 and 3 Hamming's steps of width. Of course, such a
strategy is hampered for classical computers for more wider
potentials barriers by a huge number of possible trial states.
Quantum incoherent tunneling turns out to be much more effective
because it allows to run over all local energy minima only, without
examination of all possible values of energy for all possible trial
states.

\section{\label{sec4}ELASTIC NET APPROACH TO THE TRAVELING SALESMAN
PROBLEM AND ITS QUANTIZATION}

The Traveling Salesman Problem (TSM-problem) is formulated as
follows: given positions of cities, what is the shortest tour in
which each city is visited once? It is evident that the total
number of possible tours increases exponentially with the increase
of number, $n$, of  cities: $$ N_{tours} \sim e^{C\; n},$$ and
their sequential consideration requires computing time which
increases faster than any power of $n$. Therefore problem is
considered as intractable in the classical theory of computation.
In this section we show how the problem can be solved, at least in
principle, using the quantum calculation technique. We consider
the analog  approach, because practical construction of quantum
digital computers is still confronted with serious difficulties.
Therefore, it is quite possible that quantum analog computers will
win the digital ones, despite the opposite situation in the field
of the classical computations.

Quantum analog computers for solving TSM-problem can be
constructed on the base of the Elastic Net analog algorithm
elaborated for the usual digital computers \cite{Durbin}. The
algorithm is grounded on a discrete form of the gradient equation,
$$ \frac{d \vec y_j}{dt} = - \frac{\partial F}{\partial \vec y_j}
,$$ with free energy,  $F$: $$ F= - \alpha k^2 \sum_{i} \ln
\sum_{j} e^{- | \vec x_i - \vec y_j|/2 k^2 } + \beta k \sum_{j} |
\vec y_{j+1} - \vec y_j|^2 . $$ For $\triangle t =1$ it is
possible to write: $$ \triangle \vec y_j \simeq - \frac{\partial
F}{\partial \vec y_j}, \qquad \triangle F \simeq \sum_{j}
\frac{\partial F}{\partial \vec y_j} \triangle \vec y_j \simeq -
\sum_{j}   \left( \frac{\partial F}{\partial \vec y_j} \right) ^2
< 0. $$ This means that the algorithm directs the system to a
local minimum of $F$ in accordance with the steepest descent
method. The explicit form of $F$ was devised in \cite{Durbin}
using a very clear physical picture. Positions of cities are
described by $\vec x_i$, points with coordinates $\vec y_j$ lie on
an elastic string. Each point moves under the influence of two
types of force. The first moves it towards those cities to which
it is nearest; the second pulls it towards its neighbors on the
string, acting to minimize the total string length.

The authors described an iteration procedure which consists in a
gradual decrease of the value of a parameter $K$ describing a force
and a range of interaction between $\vec x_i$ and $\vec y_j$ in such
a way that at $K$ approaching to zero the range is transformed from
a big value to  very small one and the force is strongly increased.
After applying this iteration procedure, an initial string in a form
of a small circle placed nearby the cities' centre of mass is
converted into a salesman tour of a very high quality. Comparative
analysis fulfilled in \cite{Peterson} has shown that the performance
of the Elastic Net algorithm is even of higher quality than that of
Simulated Annealing. Therefore, one can use any of them, or some
other described in \cite{Peterson}, to find a preliminary solution
to TSM-problem as an input for a further quantum
computation.



\begin{figure}
\begin{center}
\includegraphics[width=6cm,angle=90]{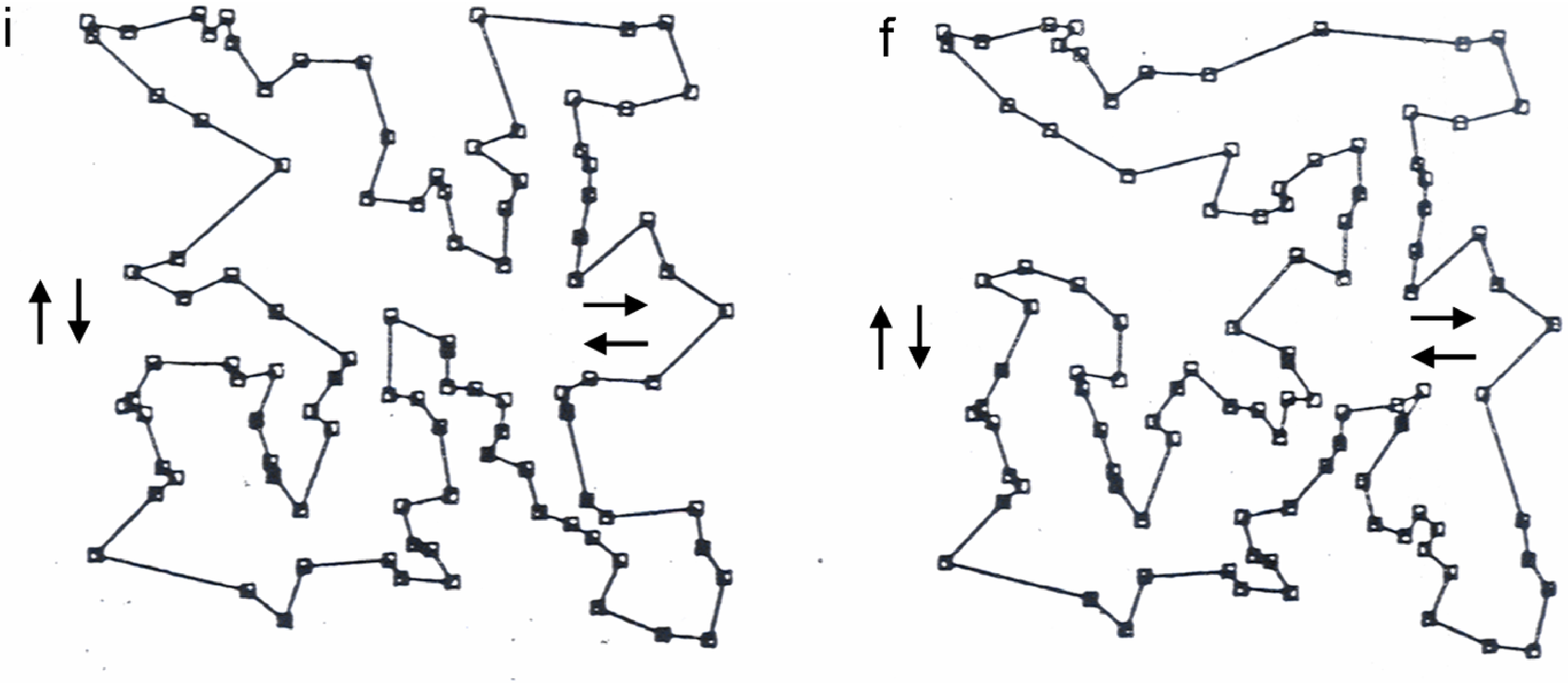}
\end{center}\vspace{-1.8cm}
\caption{\small The process of solution of TSM-problem by the domain
wall analog computer:   an input, $i$, corresponding to a local
minimum  spontaneously decays into the final state, $f$, describing
the global minimum. }\label{fig4}
\end{figure}

Elastic string can be created as a physical computation device in
the form of a mesoscopic domain wall in an antiferromagnetic thin
film at low temperature.  Tunneling of the domain walls in
ferromagnetic and antiferromagnetic insulators through potential
barriers (see, e.g., \cite{Chudnovsky}) is described in a similar
way as incoherent tunneling of particles, considered in previous
section. The role of the barriers can play defects or other singular
points in the lattice. The domain wall has a variety of dissipative
couplings to the magnons, photons, impurities, defects, and phonons.
Nevertheless, it was shown that large domain walls, containing up to
$10^{10}$ spins can behave as quantum objects at low temperatures
\cite{Stamp}. Both  theory and experiment revealed a finite
tunneling time of domain walls through potential barriers
\cite{Chudnovsky}. This gives an opportunity of devising an analog
quantum computer using the incoherent tunneling effect. The process
of solution of TSM-problem by this computer is  shown schematically
in Fig.4. Firstly, one creates an input, $i$, corresponding to some
local minimum found with the help of the classical computer, and
then, after a time, obtains a final state, $f$, as the solution.
Points in the picture denote "cities", which are some kind of
defects in the lattice that pin the domain wall to necessary
locations. They may be some implanted atoms with high value of
spins, or strongly magnetized atoms.  Energy of the domain wall per
unit of length (and, therefore, tension of the string) can be
regulated by the material properties, or by a change of the mutual
orientation of directions of spins on either side of the domain
wall. It is clear that a value of tension should be small enough to
prevent the string from a temptation to pass round some cities to
minimize its length. Such a slack string limit corresponds to the
strong shot-range interaction between the string and the cities
considered as the last step of the iteration procedure in the
classical Elastic Net analog algorithm \cite{Durbin}. Therefore,
quantum computer solution of the problem can be interpreted as a
natural continuation of the classical algorithm after taking into
account the tunneling processes. It is an easy task to examine that
at least for a square lattice with the four nearest spin-spin
interactions tension of the string is independent on its
orientation. Thus, setting aside purely engineering problems, one
may conclude that the Traveling Salesman Problem seems to be
solvable by the quantum analog computer.

\section{\label{sec5}ACKNOWLEDGMENT}

The authors are thankful to Dr. M.V. Altaisky for useful remarks.

\def\refname{

\end{document}